\documentclass[12pt]{iopart}
\usepackage{epsf}

\usepackage{graphics}
\usepackage{bm}      
\usepackage{epsfig}
\newcommand{\la}{\langle}
\newcommand{\ra}{\rangle}
\newcommand{\ua}{\uparrow}
\newcommand{\da}{\downarrow}

\begin{document}


\title{Spin-flop transition in doped antiferromagnets}

\author{N. M. R. Peres}
\address{Departamento de F\'{\i}sica
and GCEP-Centro de
F\'{\i}sica da Universidade do Minho, Campus Gualtar, P-4700-320
Braga, Portugal.}

\date{\today}

\begin{abstract}
In this paper we compute the mean field phase diagram of a doped 
antiferromagnet, in a magnetic field and
with anisotropic exchange.  
We show that at zero temperature there is a
metamagnetic transition from the antiferromagnetic configuration
along the $z$ direction to a spin-flop state. 
In the spin flop phase the system prefers a commensurate magnetic order, 
at low doping,
whereas at
larger doping the incommensurate phase is favorable. 
Contrary to
the pure Heisenberg case, the spin flop region does not span an infinite
area in the $(\Delta,h)$ plane, where $\Delta$ is the exchange anisotropy 
and  $h$ is the external magnetic field. We characterize
the magnetic and charge-transport properties
of the spin-flop phase,  computing the magnetic
susceptibility and the Drude weight. This latter quantity
presents a sudden variation as the spin-flop to paramagnet 
phase transition line is crossed. This effect could be used
as a possible source of large magneto-resistance.
Our findings may have some relevance for doped
La$_{2-\delta}$Sr$_{\delta}$CuO$_4$ in a magnetic field. 
\end{abstract}
\submitto{\JPCM}
\section{Introduction}            %
\label{s:intr}                    %

Metamagnetic  transitions are ubiquitous in Nature. 
We can find them in pure magnets, in boson systems, in
spin density wave systems, and in  doped antiferromagnets.
   
When placed in a magnetic field many magnetic materials
undergo first-order phase transitions. These materials
are called metamagnets. Two of the best studied magnetic materials
exhibiting  metamagnetic transitions are MnF$_2$ and FeCl$_2$.
\cite{shapira70,schweika02,birgeneau74}
In FeCl$_2$ there is a first-order transition from an antiferromagnetic
(AF) to a paramagnetic (P) state, and  in  MnF$_2$ 
there is a first-order transition from an AF to a spin-flop (SF)
state (see figure \ref{f:angle}, panel $a$, for 
a schematic idea on the spin configurations). 
In addition to the metamagnetic transitions above
mentioned, these materials
also undergo second-order phase transitions. The way the second-
and first-order transition lines touch each other introduces different
types of critical points. In MnF$_2$ and  FeCl$_2$ there is a 
bicritical and  a tricritical point, respectively. Some materials, as 
GdAlO$_3$, \cite{rohrer77} can present different types of critical points
depending on the orientation of the magnetic field.

From the theoretical point of view, and as early as 1936, 
N\'eel predicted a first order magnetic transition
for an anisotropic antiferromagnet in a magnetic field. 
\cite{neel} 
This metamagnetic transition was named spin-flop transition. 
Improvements of N\'eel's 
results were achieved using Green's functions and Holstein-Primakov bosons.
\cite{callen,pytte} Later, a scaling theory for  bicritical points, based on
a renormalization group analysis was introduced.\cite{fisher} 
More recently, a zero-temperature spin-flop transition 
in square and cubic lattices was studied, using exact diagonalization,
\cite{takahashi} and the SSE Monte Carlo method.\cite{yunoki}
The finite temperature study of the model phase diagram, using
exact (numerical) methods was done recently, but for the two dimensional case
only, and in the context of hard core bosons.\cite{troyer}
(The three dimensional case, at finite temperatures, was also
studied.\cite{pa})

The exact mapping between hard-core bosons with nearest neighbor
interaction and the spin 1/2 anisotropic Heisenberg model,\cite{matsuda}
permits to obtain results for both physical systems, from the
study of one of them alone.\cite{bernardet} In the boson language, the
antiferromagnetic and spin flop phases correspond to a Mott insulator phase
(a solid phase),
where the bosons are locked at the lattice sites
by the nearest neighbor repulsion, and to
the super-fluid phase, respectively. The observation of the Mott insulator 
to super-fluid transition (and vice versa) in a Bose-Einstein condensate of
$^{87}$Rb atoms confined by an optical trap,\cite{greiner} 
gave an enormous boost to the research, both theoretical
and experimental,  in this field.\cite{batroni}
The effect of disorder in the phase diagram of  the two dimensional
boson-Hubbard system has been studied as well.\cite{kim}

Also in doped magnetic materials, such as in 
La$_{2-\delta}$Sr$_{\delta}$CuO$_4$, metamagnetic transitions have been
observed.\cite{suzuki02} Both in  pure  La$_{2}$CuO$_4$ and in lightly
doped La$_{2-\delta}$Sr$_{\delta}$CuO$_4$ the spin flop transition
has its origin in the Dzyaloshinskii-Moriya interaction. Although
some theoretical study of this transition has been done in the past
for pure 
La$_{2}$CuO$_4$,\cite{tineke88} its study in doped 
La$_{2-\delta}$Sr$_{\delta}$CuO$_4$ has not been pursued to our knowledge.
The spin wave spectrum for doped La$_{2}$CuO$_4$ was 
consider Ivanov {\it et al.}.\cite{ivanov}
This motivated us to carry general studies of metamagnetic transitions
in anisotropic doped antiferromagnets. To the best of our knowledge,
the only work dealing with metamagnetic transitions in  
strong-correlated electrons was done in the context of the Hubbard model.
\cite{vollhardt97}

The paper is organized as follows: in section \ref{s:model},
we introduce the model and the possible spin phases that may
exist in the system; in section \ref{s:phase}, the zero- 
and finite-temperature phase-diagrams of the system are introduced
and discussed, together with the effect of doping on the
critical fields and on the spin-flop angle; in section \ref{s:char}
the metallic SF phase is characterized both in terms
of their magnetic and transport properties. 
 
\section{Model and mean field    %
equations}	                 %
\label{s:model}                  %

Our starting point is a simple generalization 
of the usual $t-t'-J$ model in three dimensions,
where an anisotropic exchange term, of
strength $J\Delta$, in included
\begin{eqnarray}
H= -\frac t 2 \sum_{i,\beta,\sigma}(
c^\dag_{i,\sigma}c_{i+\beta,\sigma}+H.c.)
-\frac {t'}2 \sum_{i,\beta',\sigma}(
c^\dag_{i,\sigma}c_{i+\beta',\sigma}+H.c.)\nonumber\\
+\frac J 2 \sum_{i,\beta}(\Delta S^z_iS^z_{i+\beta}+
 S^x_iS^x_{i+\beta } +  S^y_iS^y_{i+\beta})
-h\sum_{i}S^z_i -\mu\sum_{i,\sigma}
c^\dag_{i,\sigma}c_{i,\sigma}
\,.
\label{hamilt}
\end{eqnarray}
In the Hamiltonian (\ref{hamilt}), $\beta$ and $\beta'$ stand for the
lattice vectors connecting all nearest and second-nearest neighbors sites
of site $i$ in a 3D cubic lattice, respectively; the parameters
$t$ and $t'$ stand for the hopping integrals 
connecting all nearest and second-nearest neighbors sites, respectively, 
$h$ stands for the magnetic field and  $\mu$ for the chemical potential.
The density of electrons is defined as $n=N_e/N$, where $N_e$ is the total
number of electrons and $N$ is the number of lattice sites. 
Since
our calculations will be performed at electronic densities close to $n=1$
it is convenient for latter use to introduce the doping
of holes as $\delta=1-n\ll 1$.

It is a characteristic of  $t-J$ models (we are referring here to  class
of models) 
that only one electron at the most can exist at each lattice site. 
We enforce
this constraint using a slave boson representation for the 
$c_{i,\sigma}$ operators. 
\cite{coleman84}
In this representation we have
$c_{i,\sigma}=b^\dag_i f_{i,\sigma}$ and $\vec S=\frac 1 2
\sum_{\alpha',\alpha}f_{i,\alpha}^\dag\vec \sigma_{\alpha',\alpha}
f_{i,\alpha'}$, where $\vec \sigma =(\sigma_x,\sigma_y,\sigma_z)$
are the usual Pauli matrices. At each lattice site  the total number of
bosons and fermions is equal to one, and we have the
constraint $\sum_{\sigma}f_{i,\sigma}^\dag f_{i,\sigma}+ b^\dag_i b_i=1$.
In what follows, we shall consider that all bosons have condensed 
into the lowest energy state, and introduce $\la b \ra=\sqrt \delta$,
where $\delta$ corresponds, 
for zero and moderate temperatures (see finite temperature
discussion), to the doping introduced above.

\begin{figure}[ht]
\begin{center}
\epsfxsize=8cm
\epsfbox{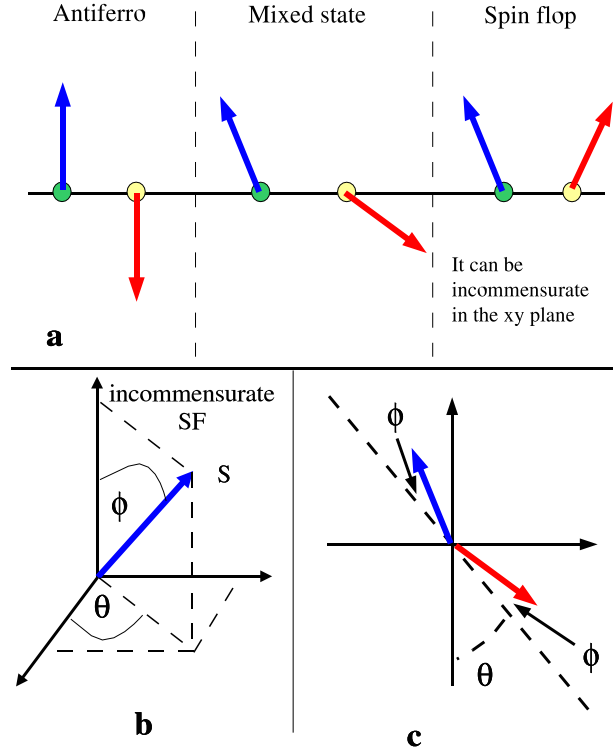}
\end{center}
\caption{
{\bf Panel a:} schematic arrangement of the spin on the two
sub lattices in the AF, mixed, and SF states.
{\bf Panel b:} definition of the angles for the spins in the spin flop
 incommensurate spiral state given by Eq. 
(\ref{flopspiral}).
{\bf Panel c:} definitions of the angles for the spins, including
 the AF, the SF, and the mixed states. The letters $S_A$ and
$S_B$ denote the spins on sub-lattices $A$ and $B$ respectively.
The spin configuration refers to the commensurate states given
by Eq. 
(\ref{aver}).}
\label{f:angle}
\end{figure}

In order to introduce a mean field Hamiltonian suitable for
the study of  metamagnetic transitions we consider two
sub-lattices, $A$ and $B$, such that in each sub-lattice the
spins are oriented as shown in Fig. \ref{f:angle}, panel $c$.
Referring to Fig. 
\ref{f:angle}, panel $c$, $\theta=\phi=0$ describes the  AF 
configuration;  $\theta=\pi/2$ and $\phi\ne 0$ describes the
SF
configuration; $\theta=\pi/2$ and $\phi=\pi/2$ describes the P configuration
(the 
term ``P configuration'' to name a state where both spins point up may
 cause some confusion, but in this context it means  the system
does not present a ferromagnetic ground state, this is, it does not
possess
a  ferromagnetic order parameter, in the sense of Landau theory
of phase transitions; the magnetic moment presented by the system
is induced by the magnetic field); 
$\theta\ne 0$ and $\phi\ne 0$ corresponds to a mixed phase, where there
is staggered magnetization in both the $z$ and $y$ directions.
The spin averages of all these configurations,
in  sub-lattices $A$ and $B$, are given by
\begin{eqnarray}
\la \vec S_{i_A} \ra &=& (0,-S_A\sin(\theta-\phi),S_A\cos(\theta-\phi))\,,\nonumber\\
\la \vec S_{i_B} \ra &=& (0,S_B\sin(\theta+\phi),-S_B\cos(\theta+\phi)).
\label{aver}
\end{eqnarray}
Note that we  consider the magnitude of the average value of the spin to be different 
at the two sub lattices. This is needed because the magnetic field induces
a certain  amount of ferrimagnetism in the system. 
The above phases are commensurate with the lattice, but it is well known
that the isotropic $t-J$ model supports spiral order in the $xy$ plane
with a momentum vector $\vec Q$ incommensurate with the lattice.  
\cite{jayaprakash89}
To account for this possibility in the presence
of magnetic field, we introduce averages
of the spin operators, presenting spiral order in the
$xy$ plane
\begin{eqnarray}
\la \vec S_i \ra = S(\sin\phi\cos\theta_i,\sin\phi\sin\theta_i,\cos\phi)\,,
\label{flopspiral}
\end{eqnarray}
where $\theta_i=\vec Q \cdot\vec R_i$, and $\phi$ represents the
angle of $\vec S$ with the $z$ axis (see figure \ref{f:angle}, panel $b$).
This phase competes with the commensurate SF order. It is also
possible that the commensurate AF order in the $z$ directions may
compete with an AF order presenting incommensurate spiral order
in the $xy$ plane 
\begin{eqnarray}
\la \vec S_{i_A} \ra &=& 
S_A(\sin\phi_A\cos\theta_i,\sin\phi_A\sin\theta_i,\cos\phi_A)\,,\\
\la \vec S_{i_B} \ra &=& 
S_B(\sin\phi_B\cos\theta_i,\sin\phi_B\sin\theta_i,-\cos\phi_B).
\label{AFinc}
\end{eqnarray}

In our study we have not found solutions  for the mixed state
and  for the AF order presenting incommensurate spiral order
in the $xy$ plane.

\subsection{Commensurate AF and SF phases} %
\label{sub:com}                            %
 
Using the averages (\ref{aver}) and introducing an Hartree-Fock
decoupling of the Hamiltonian (\ref{hamilt}) we obtain, after
Fourier transforming the operators, the following
mean field Hamiltonian
\begin{eqnarray}
H_{MF}=\sum_{k,\sigma}\epsilon_1(k)(a^\dag_{k,\sigma}b_{k,\sigma}
+b^\dag_{k,\sigma}a_{k,\sigma})+
\sum_{k,\sigma}[\epsilon_2(k)+\sigma h^z_B]a^\dag_{k,\sigma}a_{k,\sigma}+
\nonumber\\
\sum_{k,\sigma}[\epsilon_2(k)+\sigma h^z_A]b^\dag_{k,\sigma}b_{k,\sigma}+
\sum_{k,\sigma} h^x_B a^\dag_{k,\sigma}a_{k,-\sigma}+
\sum_{k,\sigma} h^x_A b^\dag_{k,\sigma}b_{k,-\sigma}\,.
\label{Hmf}
\end{eqnarray}  
where the $a^\dag_{k,\sigma}$  and the $b^\dag_{k,\sigma}$ operators 
refer to the sub-lattices $A$ and $B$, respectively,
$k$ stands for $\vec k$, 
the $k$ summation  runs over the magnetic Brillouin zone,
and
\begin{eqnarray}
\epsilon_1(k)&=&-2t\delta 
\sum_{i=x,y,z}\cos k_i\,,\hspace{1cm}\nonumber
\epsilon_2(k)=-4t'\delta \sum_{j\ne i=x,y,z}\cos k_i\cos k_j\,,
\nonumber\\
h^z_{A/B}&=& 3 J \Delta \la  S^z_{A/B} \ra -\frac h 2\,,
\hspace{1cm}\nonumber
 h^x_{A/B}= 3 J \la  S^x_{A/B} \ra\,. 
\end{eqnarray}
We point out  that in $\epsilon_1(k)$ and $\epsilon_2(k)$ 
the condensation of the bosons renormalizes the 
hoping integrals to $t\delta$ and $t'\delta$.
Since we assumed the condensation of the bosons, the constraint 
($\sum_{\sigma}f_{i,\sigma}^\dag f_{i,\sigma}+ b^\dag_i b_i=1$)
gives 
an equation for the number of particles in terms of the  doping
$\delta$
\begin{equation}
\frac 1 N \sum_{k,\sigma} (\la  a^\dag_{k,\sigma}a_{k,\sigma} \ra  +
\la  b^\dag_{k,\sigma}b_{k,\sigma} \ra) =1-\delta\,.
\end{equation} 

In the P phase there is  one
mean field parameter only: the average value of the spin $S=S_A=S_B$;
in the AF and SF phases we have the mean field parameters  $S_A$ and $S_B$, 
and
 $S$ and $\phi$, respectively. All physical quantities characterizing 
model (\ref{Hmf}) can be obtained from the associated single particle
Green's functions. 
  
\subsection{Incommensurate (spiral) SF phase}%
\label{sub:inc}                     %

Considering the spin-flop incommensurate states given by 
(\ref{flopspiral}), the mean field Hamiltonian can be cast in the form
\begin{eqnarray}
H&=&\sum_{k}[\epsilon_\uparrow(k+Q)f^\dag_{k+Q,\uparrow}f_{k+Q,\uparrow}
+\epsilon_\downarrow(k)f^\dag_{k,\downarrow}f_{k,\downarrow}]\nonumber\\
&+&\bar\Delta\sum_{k} (f^\dag_{k+Q,\uparrow}f_{k,\downarrow} +  
f^\dag_{k,\downarrow}f_{k+Q,\uparrow})+E_0\,,
\end{eqnarray}
where
\begin{eqnarray}
\epsilon_\uparrow(k+Q)&=&\epsilon_1(k+Q)+ 
\epsilon_2(k+Q) +h_z-\mu\,,\nonumber\\
h_z & =& - \frac h 2 + \frac J 2 \Delta S\cos\phi z\,,\nonumber\\
\epsilon_\downarrow(k)&=&\epsilon_1(k)+ \epsilon_2(k) -h_z-\mu\,,\nonumber\\
\bar\Delta&=&\frac J 2 S \sin\phi\gamma(Q)\,,\nonumber\\
\gamma(k)& =& 2\sum_{i=x,y,z}\cos(k_i)\,,\nonumber\\
E_0&=&-\frac J 2 \Delta S^2\cos^2\phi z N 
- \frac J 2 S^2\sin^2\phi\gamma(Q)N\,.\nonumber 
\end{eqnarray}
The mean field parameters are the amplitude $S$, the angle $\phi$
and the incommensurate momentum $\vec Q=(Q_x,Q_y,Q_z)$;
$z$ is the coordination number. The corresponding
saddle point equations are obtained from the free energy, determined from
\begin{equation}
{\cal F} = -T\sum_{k}\sum_{\alpha=\pm} \log(1+e^{-E_{\alpha}(k)/T})+
\mu(1-\delta)N+E_0\,,
\end{equation} 
with $E_{\alpha}$ equal to 
\begin{equation}
E_{\alpha} = \frac {\epsilon_\uparrow(k+Q)+\epsilon_\downarrow(k)}2
+\frac {\alpha}2 \sqrt{[\epsilon_\uparrow(k+Q)-\epsilon_\downarrow(k)]^2+
4{\bar\Delta}^2}\,\nonumber
\end{equation}
and are given in the appendix (the incommensurate momentum 
$\vec Q=(Q_x,Q_y,Q_z)$ is also determined from a saddle point equation).

We note that incommensurate spiral order in the $xy$ plane leads
to a well defined mean field theory, since  it's possible to
define a close set of equations of motion for the Green's functions.
Conversely, this is not possible for incommensurate states in the
$zx$ or $zy$ planes (in a magnetic field). 
Due to symmetry, the incommensurate states can only be of the
form $(Q,Q,Q)$, $(Q,Q,\pi)$, and $(Q,\pi,\pi)$. In Table 
(\ref{tab:incdop}) we show the effect of the doping $\delta$ on the
incommensurate $(Q,Q,Q)$ and $(Q,\pi,\pi)$  wave vectors. It is
clear from table \ref{tab:incdop} that the free energy for
the three states $(Q,Q,Q)$, $(Q,Q,\pi)$, and $(Q,\pi,\pi)$
should be  essentially the same, and therefore the value
of ${\cal F}(Q,Q,\pi)$ is not presented.
\begin{figure}[ht]
\begin{center}
\epsfig{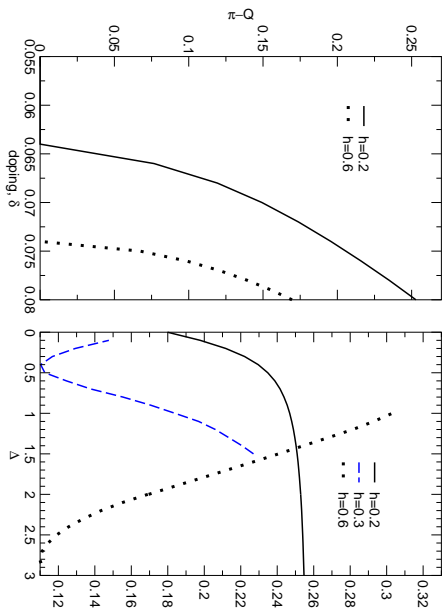}
\end{center}
\caption{Effect of doping and $\Delta$ 
on the value of the incommensurate momentum, for
$T=0.002$ (temperature in units of $t$).
The $t$, $t'$, and $J$ parameters are those of Table 
\ref{tab:incdop}, and in the left panel $h=0.2,0.6$
and $\Delta=2$; in the right panel $n=0.92$ ($\delta=0.08$) and 
$h$=0.2, 0.3, 0.6.
Please note that the scanning on $\delta$ for $h=0.2$,
corresponds to the second column of table \ref{tab:incdop}.}
\label{f:Qval}
\end{figure}

\begin{center}
\begin{table}
\begin{center}
\vspace{3mm}
\begin{tabular}{c|c|cc|cc}
\hline
$\delta$& ${\cal F}(\pi,\pi,\pi)$
        &$Q$ & ${\cal F}(Q,Q,Q)$ &  $Q$ & ${\cal F}(Q,\pi,\pi)$ \\
\hline
  0.080&  -0.17108   & 2.9996 & -0.17114&  2.8887   & -0.17115\\
  0.078&  -0.16747   &3.0066 & -0.16752&  2.9067   & -0.16753\\
  0.076&  -0.16390   &3.0142 & -0.16394&  2.9258   & -0.16394\\
  0.074&  -0.16037   &3.0228 & -0.16041&  2.9460   & -0.16040\\
  0.072&  -0.15689   &3.0327 & -0.15691&  2.9678   & -0.15691\\
  0.070&  -0.15345   &3.0448 & -0.15347&  2.9924   & -0.15347\\
  0.068&   -0.15007  &3.0617 & -0.15007& 3.0224   & -0.15007\\
  0.066&  -0.14673   &3.0891 & -0.14673&  3.0649   & -0.14673\\
  0.064&   -0.14345  &  $\pi$& -0.14345 & $\pi$    & -0.14345\\
\hline
\end{tabular}
\end{center} 
\caption{Free energy values for incommensurate momentum $Q$ as function of
density, for $T=0.002$, $h=0.2$, $\Delta=2$. Here and in the 
remaining figures we have used $t=1$, $t'=0.1$ and $J=0.1$. 
These results have been obtained in a lattice of 100$\times$100
$\times$100.}
\label{tab:incdop}
\end{table}
\end{center}

It is clear from table \ref{tab:incdop} that the incommensurate spiral
phase has a lower free energy for moderate doping. As the doping is reduced
the system finds the commensurate phase energetically favorable. 
\cite{note1,note2}
A detailed
evolution, as function of doping and $\Delta$, of the incommensurate spiral
phase for the momentum  $(Q,\pi,\pi)$ is presented in figure 
\ref{f:Qval}. In this figure we present results for different 
values of the magnetic field ($h=$0.2, 0.3, 0.6). 
The behavior of $Q$
with $\Delta$ is seen to be non monotonous. For a given value
of $h$, $\pi-Q$ may present a minimum ($Q$ maximum) for a given value
of $\Delta$. If $h$ is large (for example, $h=0.6$), such that
as $\Delta$ is reduced the SP to P transition-line is crossed, we 
see $Q$ tends to flow away from the commensurate $\pi$ value.
In the other cases (for example, $h=0.3$), $Q$   approaches the
commensurate value first ($\pi-Q$ is reduced) , reaches the  value
where $\pi-Q$ is minimum, 
and starts to deviate again from the commensurate value ($\pi-Q$ increases).
(In the P phase there is no meaning for the incommensurate
or commensurate states.)
\section{Phase diagram at zero   %
 and finite temperatures}	 %
\label{s:phase}                  %

\begin{figure}
\begin{center}
\epsfig{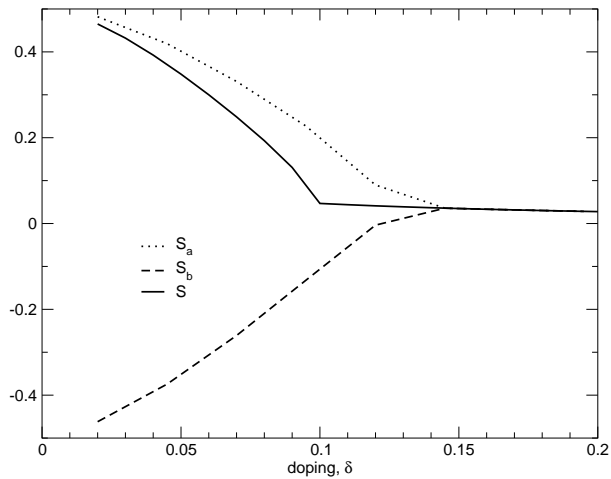}
\end{center}
\caption{Effect of doping on the magnetic structure 
at zero temperature. The $t$, $t'$, and $J$ parameters are those of Table 
\ref{tab:incdop}, and $\Delta=1.2$ and $h$=0.1. For these values
of the parameters the AF phase is the   more stable phase.} 
\label{f:dopping}
\end{figure}

At $T=0$ and $\delta=0$ it is a simple exercise to obtain an 
analytical solution for the phase diagram
of the system. There is no spiral order and 
the paramagnetic ($E_P$), antiferromagnetic ($E_{AF}$), and
spin flop $(E_{SF})$ energies are given (for spin 1/2) by
\begin{equation}
E_{P}=\frac 1 4 Jz\Delta- h\,,  \hspace{0.5cm}
E_{AF}=-\frac {Jz\Delta}4\,,
\label{e1}
\end{equation}
and
\begin{equation}
\hspace{1cm} E_{SF}=-\frac {Jz}4 
-\frac {h^2}{Jz(1+\Delta)}\,,
\label{e2}
\end{equation}
respectively. At zero temperature the average value of $S$, $S_A$, and $S_B$
is 1/2. 
From results (\ref{e1}) and (\ref{e2}) the zero temperature
phase diagram of the system can be obtained. The line 
separating the AF and SF phases is given by  $h=zJ\sqrt{\Delta^2-1}/2$,
and the  SF and  P phases are separated by the line
$h=zJ(\Delta +1)/2$. 
When we dope the antiferromagnet, there is a reduction of the 
average values of $S$, $S_A$, and $S_B$ relative to 1/2, even at zero
temperature, as can be seen from figure \ref{f:dopping}. This behavior
is well known for the $2D$ isotropic $t-t'-J$ model, 
and has been used to explain,
at the mean field level, the reduction of the N\'eel temperature 
in high-temperature superconductors, upon doping. Here, the presence
of the magnetic field combined with doping  forces the AF phase
to a ferrimagnetic phase, where the average value  of spin  in the 
two sub-lattices is not the same even at zero temperature. This is in contrast
to the pure Heisenberg case, where $S_A=S_B=1/2$ at zero temperature
in the presence of $h$.  Upon increasing doping, the average
values of $S$, $S_A$, and $S_B$ are reduced to the paramagnetic value,
fixed by the magnetic field. Above a given value of $\delta$, there
is no magnetic order in the ground state of the system, and its behavior
it that of a collection of fully polarized independent electrons.

This reduction of $S$, $S_A$, and $S_B$ together with the combined effect of
$\Delta$ and $h$ has consequences for the zero temperature phase diagram
of the doped antiferromagnet, when compared with the pure Heisenberg 
case discussed above.  The picture
is the following: doping introduces holes here and there in the lattice
and as consequence some of the magnetic interactions due to the Heisenberg term
cannot be fulfilled; as a consequence, and for a given $\Delta >1$, as the
magnetic field increases, the system finds favorable to take advantage
of the Zeeman energy and therefore the first order AF-SF
transition should occur at a lower value of $h$; for the same reason,
the SF phase cannot fulfill all the antiferromagnetic 
interactions in the $xy$ plane, and therefore the second order SF-P
transition should also occur at lower values of
$h$.
\begin{figure}
\begin{center}
\epsfig{figure=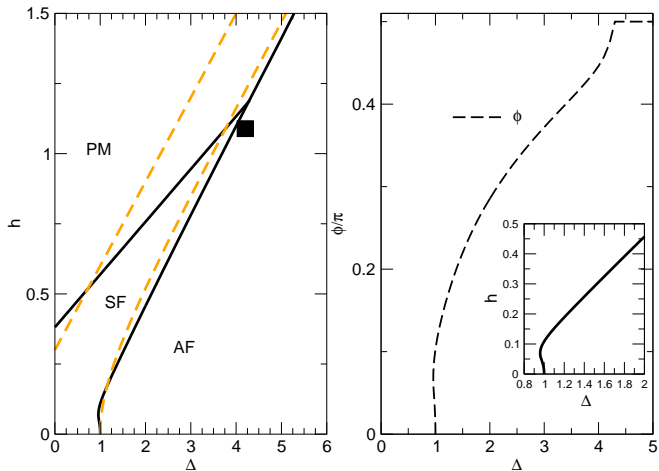,width=8cm,angle=-90}
\end{center}
\caption{{\bf Left panel:} zero temperature phase diagram of 
the 3D $t-t'-J$ model,
for $1-\delta=n=0.92$. The $t$, $t'$, and $J$ parameters are those of Table 
\ref{tab:incdop}. The black box marks the point where the first-order
AF to SF and the second-order SF to PM transition lines meet. For values of
$\Delta$ above the black box there is a AF to PM second-order
transition line only. The pure Heisenberg phase diagram is 
represented by the dashed lines. The full lines represent
the phase diagram for the doped case. 
{\bf Right panel:}  variation of the SF angle
$\phi$, in units of $\pi$,
along the AF to SF transition-line. The inset shows the phase diagram
close to $\Delta=1$. A minute reentrant behaviour is seen.} 
\label{f:phase0}
\end{figure}

This picture is confirmed by the phase diagram of figure 
\ref{f:phase0}. In this figure, and for $\Delta$ roughly in the range 
$1<\Delta <4$ we see that both the first order AF-SF and
the second order SF-P transition lines are pulled down to lower values
of $h$ relative to the pure Heisenberg case. Furthermore, and in contrast
to the pure Heisenberg case, the SF phase does not span an infinite area
in the ($\Delta,h$) plane. There is a point where the second order 
SP-P line meets the  first order AF-SF line (represented by
a black square in figure \ref{f:phase0}). From this point on we are left with
an AF-P second order transition;  below this line the spins in the $AF$ phase
are fully aligned with the field, but have different magnitudes for the two sub-lattices. 
 For values  of $\Delta<1$ the system finds preferable to
be in de SF phase. We note that in figure  \ref{f:phase0}
the spin flop phase is of incommensurate type, in agreement
with the results of figure \ref{f:Qval}. In fact using figure 
\ref{f:Qval} in connection with the phase diagram of figure \ref{f:phase0}
it is possible to see the evolution of the incommensurate
momentum $Q$ over the phase diagram.

\begin{figure}
\begin{center}
\epsfig{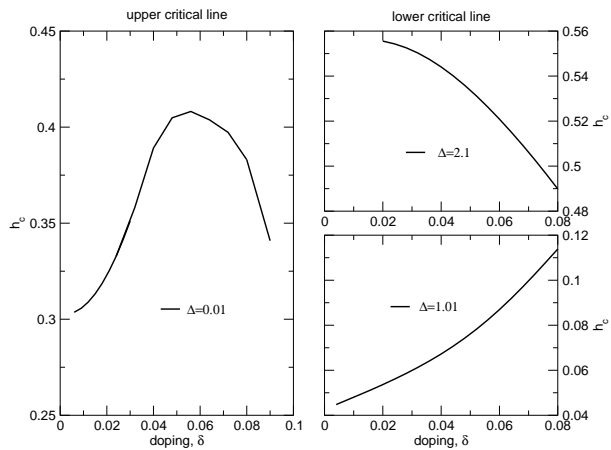}
\end{center}
\caption{Effect of doping on the critical field $h_c$.
In the left panel we show the effect of $\delta$
on the SF-P transition, close to the $\Delta=0$ case. The
right panel shows the value of the critical field 
$h_c$ for the AF-SF transition for two values of $\Delta$.
The $t$, $t'$, and $J$ parameters are those of Table 
\ref{tab:incdop}.
} 
\label{f:dophc}
\end{figure}

We can now ask, how does the transition lines change with the doping?
This question has experimental relevance in connection with measured
spin flop fields in La$_{2-x}$Sr$_x$CuO$_4$,\cite{suzuki02} where
it was found that the critical field for the SF transition
is reduced upon doping. In figure \ref{f:dophc} we show
both the effect of $\delta$ on the SF-P line, for small
values of $\Delta$ (left panel), and on the $AF-SF$ for
two values of $\Delta$. We see that for values of the exchange
$\Delta= 2$ there is a reduction of $h_c$ upon doping
(we remark this behavior is not restricted to the single
value $\Delta=2$, but it exists over a finite range of $\Delta$ values).
On the other hand, the effect for values such that  $\Delta\sim 1$
is to introduce an increase of $h_c$. Although our calculation cannot
be extended quantitatively to  La$_{2-x}$Sr$_x$CuO$_4$ mainly 
 because the spin flop transition in this material
 is due to the Dzyaloshinskii-Moriya
interaction we do not expect
qualitative changes in what respects the behavior of
$h_c$ with doping.\cite{eduardo02} That is, it is possible to account
at the mean field level for the decrease of $h_c$ with doping, independently of the details
of the interaction, as long as a AF-SF transition exists.      

\begin{figure}
\begin{center}
\epsfig{figure=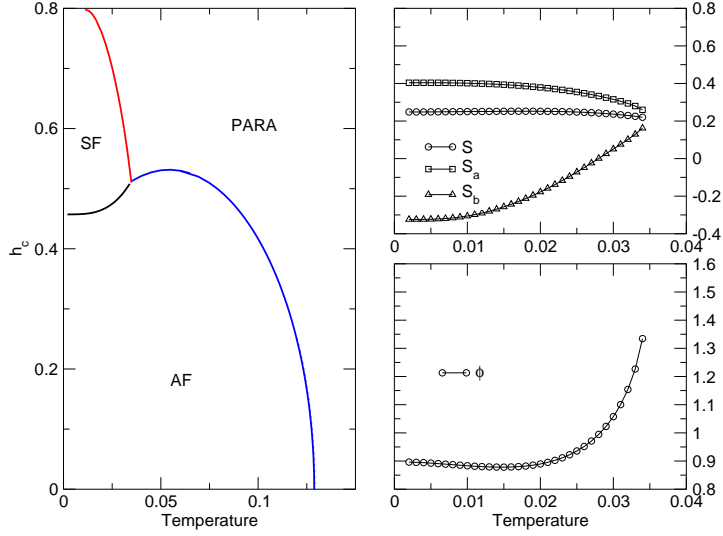,width=8cm,angle=-90}
\end{center}
\caption{Finite temperature phase diagram of the 3D $t-t'-J$ model,
for $\Delta$=2 and $1-\delta=n=0.92$.
The $t$, $t'$, and $J$ parameters are those of Table 
\ref{tab:incdop}. The left panel shows the phase
diagram, and the right panels show $S$, $S_A$, $S_B$, and $\phi$
as function of $T$ along the AF to SP first-order transition line.} 
\label{f:phase1}
\end{figure}

Let us see now
how  the zero-temperature picture evolves when 
we extend the analysis to  finite temperatures. 
Since the Bose-Einstein condensation temperature is of the order of the doping,
$T_{BE}\sim \delta t$, we can  draw a finite temperature phase diagram 
using the same formulation we used for zero $T$ as long as $T<T_{BE}$.
We have stayed in this regime.
As for the pure 
Heisenberg case, we obtain a bicritical point where the first order
AF-SF transition line meet the two second order lines, describing
AF-P and the SF-P transitions. Comparing with the pure case,
the main changes are: (i) the bicritical point $(T_b,h_b)$ moves
to lower temperatures; (ii) the SF region shrinks as we dope
the system up to a point where only the AF and the P phases remain
(the area of the AF zone is also reduced).
The disappearance of the SF phase with doping before the AF phase,
is related to the zero temperature dependence
of the average value $S$ upon doping that is seen in figure \ref{f:dopping},
where the $S$ attains its paramagnetic value before
the $S_A$ and $S_B$ do so. The aspects discussed  above are illuminated
in figure \ref{f:phase1}, where in the left panel the phase
diagram is depicted, and in the right one we plot the dependence of the
average values of $S$, $S_A$, and $S_B$,
and the angle $\phi$ in the SF phase 
along the first order AF-SF transition line. We see that close to
the bicritical point, and in a reduced range of temperatures,
the angle $\phi$ has  a fast variation. Therefore, for an experiment
probing the AF-SF transition with temperature, at different fields close
to the bicritical point we may see a large or small jump in 
the magnetization, depending on the field strength. Such a behavior 
could be easily observed performing magnetization measurements
of the type presented for
La$_{2-x}$Sr$_x$CuO$_4$, \cite{suzuki02} but with the
magnetic field applied along the  Cu0$_4$ planes.
Again, our discussion of the relation
 between our findings and the experimental
results in  La$_{2-x}$Sr$_x$CuO$_4$ applies with the limitations
discussed above.

\section{Characterization of the %
spin flop phase}	         %
\label{s:char}                   %

Let us characterize the metallic SF phase. We consider 
both magnetic and charge transport properties. We first compute
the magnetic longitudinal and transverse susceptibility.
After we study the optical conductivity, computing both
the Drude weight and the regular part of the conductivity.
\subsection{Magnetic             %
susceptibility}	                 %
\label{sub:susceptibility}       %

The dynamic magnetic susceptibility, $\chi_{\alpha,\beta}(q,i\omega_n)$,
is defined as
\begin{equation}
\chi_{\alpha,\beta}(q,i\omega_n)=\int_0^\beta d\,\tau e^{-i\omega_n\tau}
\la T S_\alpha(q,\tau)
S_\beta(-q,0)\ra\,,
\label{chidef}
\end{equation}
where
\begin{equation}
S_\mu(q)=\frac 1 {2N}\sum_{k,\alpha,\beta}f^\dag_{k+q,\alpha}
\sigma_\mu^{(\alpha,\beta)}f_{k,\alpha}\,.
\end{equation}

In the  SF phase we can define four types of Green's functions,
generically written as 
\begin{equation}
G_{\alpha,\beta}(k,p,\tau) = -\la T f_{k,\alpha}(\tau)
f^\dag_{p,\beta}(0)\ra\,.
\label{greendef}
\end{equation}
For each case the results in the Matsubara representation are
\begin{eqnarray}
G_{\ua,\ua}(k,p,i\omega_n) &=& 
\frac {\delta_{k,p}[i\omega_n-\epsilon_\da(k-Q)]}
{[i\omega_n-\epsilon_\da(k-Q)][i\omega_n-\epsilon_\ua(k)]-{\bar\Delta}^2}\,,
\nonumber\\
G_{\da,\ua}(k,p,i\omega_n) &=& 
\frac {\bar\Delta G_{\ua,\ua}(k+Q,p,i\omega_n)}
{i\omega_n-\epsilon_\da(k)}\,,\nonumber\\
G_{\da,\da}(k,p,i\omega_n) &=& 
\frac {\delta_{k,p}[i\omega_n-\epsilon_\ua(k+Q)]}
{[i\omega_n-\epsilon_\da(k)][i\omega_n-\epsilon_\ua(k+Q)]-{\bar\Delta}^2}\,,
\nonumber\\
G_{\ua,\da}(k,p,i\omega_n) &=& 
\frac {\bar\Delta G_{\da,\da}(k-Q,p,i\omega_n)}
{i\omega_n-\epsilon_\ua(k)}\,.\nonumber
\end{eqnarray}
Since $G_{\ua,\ua}(k,p,i\omega_n)$ and $G_{\da,\da}(k,p,i\omega_n)$
are diagonal in momentum space, the only non-zero off-diagonal Green's
functions in spin space are  $G_{\da,\ua}(p-Q,p,i\omega_n)$ and 
$G_{\ua,\da}(p+Q,p,i\omega_n)$.

At the mean field level,
the longitudinal magnetic susceptibility, $\chi_{z,z}$, and
the transverse magnetic susceptibility, $\chi_{-,+}$,  are obtained using
the definition (\ref{chidef}) combined with the Green's
functions (\ref{greendef}). The general form of the susceptibility
is 
\begin{eqnarray}
\chi_{\alpha,\beta}(q,i\omega_n)=-\frac 1 N\sum_p \sum_{i,l=1}^2
T[E_j(p+q),E_l(p),i\omega_n]\nonumber\\ 
\times [M_{j,l,\alpha,\beta }(p,q)]^2\,,
\end{eqnarray}
where 
\begin{equation}
T[E_j(p+q),E_l(p),i\omega_n]
=
\frac {f[E_j(p+q)]-f[E_l(p)]}{i\omega_n+E_j(p+q)-E_l(p)}\,,
\end{equation}
and
\begin{equation}
M_{j,l,z,z}(p,q)=\frac 1 2
[R_{\ua,j}(p+q)R_{\ua,l}(p)-R_{\da,j}(p+q)R_{\da,l}(p)]\,.
\end{equation}
\begin{equation}
M_{j,l,-,+}(p,q)=R_{\da,j}(p+q)R_{\ua,k}(p-Q)\,,
\end{equation}
Here $f(x)=(1+e^{x/T})^{-1}$ 
and the $R_{\alpha,j}$ factors are given in \ref{ap:green}.
\begin{figure}
\begin{center}
\epsfxsize=8cm
\epsfbox{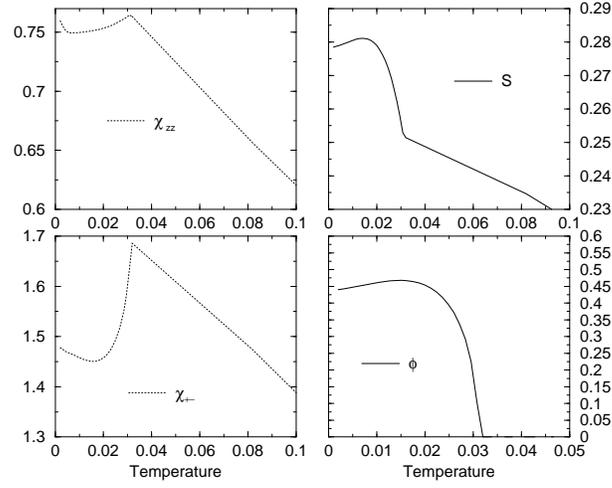}
\end{center}
\caption{Spin susceptibility $\chi_{z,z}(0,0)$ and 
$\chi_{-,+}(0,0)$ at finite 
temperature in the spin-flop region of the
phase diagram  
for $\Delta$=2, $h=0.6$ and $n=0.92$.
The $t$, $t'$, and $J$ parameters are those of Table 
\ref{tab:incdop}.} 
\label{f:sus}
\end{figure}

The behavior of $\chi_{z,z}(0,0)$ and $\chi_{-,+}(0,0)$
is presented in figure (\ref{f:sus}).
Above the SF-P transition line defined by the phase
diagram of figure \ref{f:phase1}, the behavior of 
the susceptibility is that of a paramagnet
in a magnetic field. As $h\rightarrow 0$, $\chi_{+-}=2\chi_{zz}$.
Below the transition temperature, there is a sudden drop
in the magnetic susceptibility, followed by the same abrupt
behavior of the SP angle $\phi$. 
As the temperature is further reduced
the susceptibility shows an up turn
at low temperature, which  
is typical of the susceptibility 
of metallic antiferromagnets, as is the case of Pt$_3$Fe.
\cite{crangle}
\subsection{Optical conductivity}%
\label{sub:optical}              %

The response of the system to an electromagnetic
field is obtained from the optical conductivity. This quantity
in defined as \cite{mahan}

\begin{equation}
\sigma_{xx}(\vec q,\omega)=
\frac 1 N\frac{\la K_{xx}\ra +
\Lambda_{xx}(\vec q,\omega)}
{i(\omega+i0^+)}\,,
\end{equation}
where $\Lambda_{xx}(\vec q,\omega)$ is the retarded 
 current-current correlation function, obtained from the corresponding Matsubara
correlation function
\begin{equation}
\Lambda_{xx}(\vec q,i\omega_n)=\int_0^\beta d\tau e^{i\omega_n\tau}
\la T_\tau j^p_x(\vec q,\tau)j^p_x(-\vec q,0)\ra \,,
\label{LB}
\end{equation}
and $K_{xx}$
is given by
\begin{equation}
K_{xx}= \sum_{k}e(k+Q)f^{\dag}_{\ua,k+Q}
f_{\ua,k+Q} + e(k)f^{\dag}_{\da,k}
f_{\da,k}\,.
\end{equation}
Here $e(k)=-2t\delta \cos q_x -4t'\delta \cos q_x (\cos q_y+\cos q_z) $
and $j^p_x(\vec q)$ is a Fourier component of the current operator.
\cite{mahan}
Writing  $\sigma(\vec q,\omega)=\sigma'(\vec q,\omega) + 
i\sigma''(\vec q,\omega)$, that is, separating the real and imaginary parts, 
the real part reads
\begin{eqnarray}
\sigma'(\vec q,\omega)&=&
- \frac {\pi} N \delta(\omega)[
\la K_{xx}\ra+ \Lambda_{xx}'
(\vec q,\omega)]
+ 
\frac {\Lambda_{xx}''(\vec q,\omega)}{N\omega}
\nonumber\\
&=&\pi \delta(\omega)D
+\sigma_{reg}(\vec q,\omega)\,,
\label{realsigma}
\end{eqnarray} 
where $D$ is the charge stiffness or Drude weight, given by
\begin{equation}
D=-\frac 1 N [\la K_{xx}\ra + \Lambda'_{xx}(0,0)]\,. 
\end{equation}  
The zero momentum conductivity is given by
\begin{eqnarray}
\sigma_{reg}(\omega)&=&\frac {\Lambda_{xx}''(0,\omega)}{N\omega}
=\pi \sum_{p}\sum_{m\ne j=1}^2 M_{m,j}\delta (\omega+E_j-E_m)
\,,
\end{eqnarray}
with 
\begin{eqnarray}
M_{m,j}&=&
[f(E_j)-f(E_m)][
j^2(p+Q)R^2_{\ua,m}R^2_{\ua,j} \nonumber\\
&+&2j(p+Q)j(p)+ R_{\ua,m}R_{\ua,j}R_{\da,m}R_{\da,j} 
\nonumber\\
&+& j^2(p)R^2_{\da,m}R^2_{\da,j}]\,,
\end{eqnarray}
where $j(k)=2t\delta \sin q_x +4t'\delta \sin q_x (\cos q_y+\cos q_z) $.

\begin{figure}
\begin{center}
\epsfxsize=8cm
\epsfbox{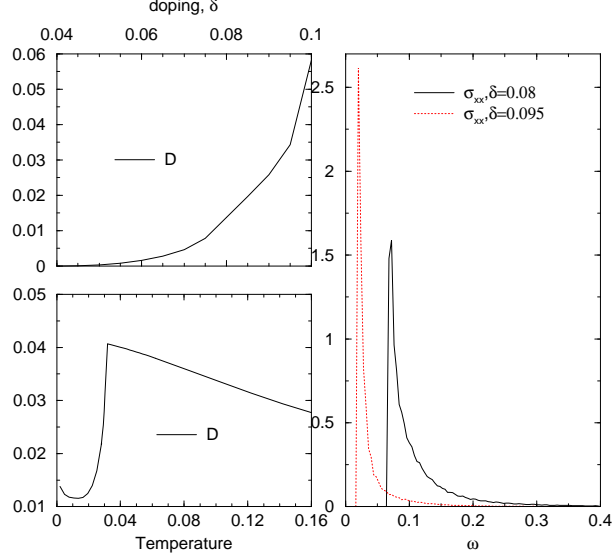}
\end{center}
\caption{Optical conductivity and Drude weight of the 3D $t-t'-J$ model,
for $\Delta$=2 and $n=0.92$, $h$=0.6.
The $t$, $t'$, and $J$ parameters are those of Table 
\ref{tab:incdop}.
In the left panels we show the effect of the doping on the 
zero temperature Drude weight $D$ and the effect 
of the temperature on $D$ as the SF-P transition line is
crossed. In the right panel, the regular part  $\sigma_{reg}$
of the optical conductivity is depicted.} 
\label{f:optica}
\end{figure}

A finite Drude weight establishes an infinite $d.c.$ conductivity.
If the system is an insulator $D$ is zero. On the other hand,
$\sigma_{reg}(\omega)$, establishes the absorption of 
finite frequency light 
 by the system. In figure \ref{f:optica}
we show the effect of the doping and of the temperature
on $D$, as well as the effect of the doping on $\sigma_{reg}(\omega)$
at zero temperature.  At zero temperature we realize that the effect of
increasing $\delta$ is two fold: 
(i) it increases $D$ (more carriers available);  (ii) it shifts 
$\sigma_{reg}(\omega)$ to lower values in frequency, because the
gap decreases with $\delta$. When $\delta\rightarrow 0$, both 
$D$ and $\sigma_{reg}(\omega)$ vanish. We expect this result to hold
beyond our mean field analysis. \cite{cote95}
At finite temperature and constant $\delta$, there are two situations.
Above the SF-P transition the Drude weight exhausts the sum rule
obeyed by $\sigma(\omega)$, and therefore there is no finite 
energy absorption. When the SF-P transition line is crossed, there
is a reduction of $D$, with a transfer of spectral
weight to finite energies. The diminishing of $D$ is quite
abrupt in a small range of temperatures, and this fact  opens
the possibility of having a large magneto-resistance
in these systems, when scattering from impurities is
included.
\section{Conclusions}            %
                                 %

In conclusion, we have, in this article, described the mean field theory of 
spin-flop transitions in a doped antiferromagnets. Further
we have characterized the spin-flop phase from the point of view
of its magnetic and  charge-transport properties. We have shown the
AF phase is characterized by a field-induced ferrimagnetic ground state,
at odds with the pure Heisenberg case,
and that a SF phase exists in a finite range of $\Delta$ and $h$. 
The Drude weight of the system was shown to present an abrupt decrease in 
a reduced range of temperatures when the system enters the spin-flop phase,
and a similar behavior occurs with the magnetic susceptibility. 
We have also studied the behavior of the critical lines with the doping, and
found that depending on the value of $\Delta$ the critical field associated
with the AF to SF transition-line may increase ($\Delta\sim 1$)
or decrease ($\Delta>1$) with the doping $\delta$. 
We have also shown that the behavior of the critical field associated with
the SF to P transition is non monotonous with $\delta$. 
Although, the full magnetic behavior of doped La$_2$CuO$_4$ in a magnetic field
cannot  be quantitatively described by an antiferromagnet 
with an anisotropic exchange, since
the origin of the spin-flop phase in this material 
is due to the Dzyaloshinskii-Moriya interaction,
we believe that some of the above qualitative 
findings still apply. The main reason, we
believe, is the fact that some physical properties will not depend so much
on the details of the interaction but on the existence of a physical mechanism
allowing for a spin flop transition. Only this reasoning may explain why
the  upper right panel of Fig. 6 exhibits a decreasing of the lower critical field
with doping, as  observed in doped La$_2$CuO$_4$. Also the existence of an incommensurate
spin flop phase will not depend on the detail of the interaction but only on the competition
between the kinetic and interaction energies. The reason why an incommensurate spin flop
state is not observed in pure La$_2$CuO$_4$ is precisely the absence of the kinetic 
energy term. On the contrary,  details as  the line borders separating the different
phases, the values of the incommensurate momentum and the amount of reduction of the critical field
with doping will certainly be interaction dependent. 
\ack                             %
                                 %
The author 
thanks J. M. B. Lopes dos Santos 
for many discussions about critical points and phase transitions, and
P. D. Sacramento and J. L. Ribeiro for carefully reading the manuscript and
suggestions.

\appendix
\section{Green's function          %
decomposition, mean field equations %
  and susceptibility}              %
\label{ap:green}                   %

In the spin flop phase all the Green's functions
can be cast in the form
\begin{equation}
G_{\alpha,\beta}(p,i\omega_n)= \sum_{j=1}^2
\frac {R_{\alpha,j}R_{\beta,j}}{i\omega_n-E_j}\,.
\end{equation}
In particular we have that
\begin{eqnarray}
G_{\ua,\ua}(p+Q,i\omega_n)&= &\sum_{j=1}^2
\frac {[R_{\ua,j}(p)]^2}{i\omega_n-E_j(p)}\,,\nonumber\\
G_{\da,\da}(p,i\omega_n)&=&\sum_{j=1}^2
\frac {[R_{\da,j}(p)]^2}{i\omega_n-E_j(p)} \,,\nonumber\\
G_{\da,\ua}(p,p+Q,i\omega_n)&= &
\sum_{j=1}^2
\frac {R_{\da,j}(p)R_{\ua,j}(p)}{i\omega_n-E_j(p)}\,,\nonumber\\
G_{\ua,\da}(p+Q,p,i\omega_n)&= &
\sum_{j=1}^2
\frac {R_{\da,j}(p)R_{\ua,j}(p)}{i\omega_n-E_j(p)}\,,\nonumber
\end{eqnarray}
where the coherence factors are
\begin{eqnarray}
[R_{\ua,1}(p)]^2&=&[R_{\da,2}(p)]^2=\frac 1 2 \left(
1+\frac {\epsilon_\da(p)-\epsilon_\ua(p+Q)}{\sqrt{\alpha(p)}}
\right)\,,\nonumber\\
1&=&\sum_{j=1}^2[R_{\alpha,j}(p)]^2\,,\nonumber
\end{eqnarray}
\begin{equation}
R_{\da,1}(p)R_{\ua,1}(p)=
-R_{\da,2}(p)R_{\ua,2}(p)=-\frac {\bar \Delta}{\sqrt{\alpha(p)}}
\,,\nonumber
\end{equation}
with 
$\alpha(p)= [\lambda(p)]^2+
4{\bar\Delta}^2$ and 
$\lambda(p)=\epsilon_\uparrow(p+Q)-\epsilon_\downarrow(p)$.

The mean field equations can be expressed in terms
of the coherence factors as 
\begin{eqnarray}
1-\delta =\frac 1 N \sum_{k}\sum_{j=1}^2(R_{\ua,j}^2+R^2_{\da,j})f(E_j)\,,\\
S\sin(2\phi)[\Delta z-\gamma(Q)]=-\gamma(Q)\cos\phi\frac 1 N
\sum_k\sum_{j=1}^22R_{\ua,j}R_{\da,j}f(E_j)+\nonumber \\
\Delta z \sin\phi
\frac 1 N \sum_{k}\sum_{j=1}^2(R_{\ua,j}^2-R^2_{\da,j})f(E_j)\,,\\
S[z\Delta \cos^2\phi+\gamma(Q)\sin^2\phi]=
\Delta z \cos\phi\frac 1 {2N}
\sum_{k}\sum_{j=1}^2(R_{\ua,j}^2-R^2_{\da,j})f(E_j)
+\nonumber \\
\gamma(Q) \sin\phi\frac 1 {2N}
\sum_k\sum_{j=1}^22R_{\ua,j}R_{\da,j}f(E_j)
 \,.
\end{eqnarray}
The above equations hold for both the commensurate and incommensurate
cases. In the incommensurate case an additional equation is required
for the determination of $Q_j$ ($j=x,y,z$)-- the incommensurate momentum
value. This equation reads 
\begin{eqnarray}
\frac {\partial {\cal F}}{\partial Q_j}=0
\Leftrightarrow  \frac 1 N \sum_{k}\sum_{\alpha=\pm}
\frac {\partial E_{\alpha}}{\partial Q_j}f( E_{\alpha})+
JS^2\sin^2\phi\sin Q_j=0\,,
\end{eqnarray}
the partial derivatives of the quasi-particle dispersions 
$E_{\alpha}$ are 
straightforward but give long equations  we omit here.

The full expression for the $\chi_{zz}(0,0)$ susceptibility 
useful for any value of the doping $\delta$ is
given by
\begin{eqnarray}
\chi_{zz}(0,0) &=& -\sum_{k,\alpha=\pm}
\alpha
f(E_\alpha)
\frac {2{\bar \Delta}^2}
{[\alpha(k)]^{3/2}}\nonumber\\
&+&
\sum_{k,\alpha=\pm}\frac 1 4\frac 
{[\lambda(k)]^2
f(E_\alpha)f(-E_\alpha)}
{T\alpha(k)}\,.
\end{eqnarray}



\begin{thebibliography}{99}
\bibitem{note2}
The smallness of the energy difference between the
commensurate and incommensurate states raises
the question about what state will actually be observed in
a real material characterized by such a small energy difference.
 In our opinion,  
such a material would
present coexistence
of macroscopic domains where some of the spins 
either order in a commensurate or
incommensurate state (as in an ordinary ferromagnet
we have different ferromagnetic domains ). 
However we stress that the observation 
of the spin-flop state, either in a commensurate or 
incommensurate state, is not in question.
\end{thebibliography}
\end{document}